\documentclass[aps,prd,showpacs,nobibnotes,twocolumn,preprintnumbers,
nobalancelastpage,superscriptaddress]{revtex4}
\usepackage{amssymb}
\usepackage{graphicx}
\usepackage{latexsym}
\usepackage{dcolumn}
\usepackage{bm}
\input{epsf}
\newcommand{\be}{\begin{equation}}
\newcommand{\ee}{\end{equation}}
\newcommand{\ba}{\begin{eqnarray}}
\newcommand{\ea}{\end{eqnarray}}
\newcommand{\no}{\nonumber}
\newcommand{\bfi}{\begin{figure}
\epsfxsize=8cm
\epsffile}
\newcommand{\bfig}{\begin{figure*}
\epsfxsize=15cm
\epsffile}
\newcommand{\efi}{\end{figure}}
\newcommand{\efig}{\end{figure*}}
\newcommand{\bi}{\begin{itemize}}
\newcommand{\ei}{\end{itemize}}

\newcommand{\etal}{{\it et al.}}
\newcommand{\mpch}{h^{-1} {\rm Mpc}}

\newcommand{\dif}{\mathrm{d}}

\newcommand{\bibmnras}{Mon. Not. R. Astron. Soc. }
\newcommand{\bibapj}{Astrophys. J. }
\newcommand{\bibapjl}{Astrophys. J. Let. }

\newcommand{\bibprd}{Phys. Rev. D } 
\newcommand{\bibprl}{Phys. Rev. Lett. }

\begin{document}

\title{Gaussianizing the non-Gaussian lensing convergence field II: the
  applicability to noisy data}
\author{Yu Yu}
\email{yuyu22@shao.ac.cn}
\affiliation{Key laboratory for research in galaxies and cosmology, Shanghai Astronomical Observatory, Chinese Academy of
 Science, 80 Nandan Road, Shanghai, China, 200030}
\author{Pengjie Zhang, Weipeng Lin}
\affiliation{Key laboratory for research in galaxies and cosmology, Shanghai Astronomical Observatory, Chinese Academy of
  Science, 80 Nandan Road, Shanghai, China, 200030}
\author{Weiguang Cui}
\affiliation{Key laboratory for research in galaxies and cosmology, Shanghai Astronomical Observatory, Chinese Academy of
  Science, 80 Nandan Road, Shanghai, China, 200030}
\affiliation{Astronomy Unit, Department of Physics, University of Trieste, Tiepolo 11, I-34131 Trieste, Italy}
\author{James N. Fry}
\affiliation{Department of Physics, University of Florida, Gainesville Florida 32611-8440, USA}
\begin{abstract}
In paper I (Yu \etal \cite{Yu11}), we showed through N-body simulation that a local monotonic Gaussian transformation
 can significantly reduce non-Gaussianity in a noise-free lensing convergence field.
This makes the Gaussianization a promising theoretical tool to understand high-order lensing statistics.
Here we present a study of its applicability in lensing data analysis,
 in particular when shape measurement noise is presented in lensing convergence maps.
(i) We find that shape measurement noise significantly degrades the Gaussianization performance
 and the degradation increases for shallower surveys.
(ii) The Wiener filter is efficient in reducing the impact of shape measurement noise.
The Gaussianization of the Wiener-filtered lensing maps is able to suppress skewness, kurtosis,
 and the 5th- and 6th-order cumulants by a factor of 10 or more.
It also works efficiently to reduce the bispectrum to zero.

\end{abstract}
\pacs{98.80.-k; 98.65.Dx; 98.62.Ve; 98.62.Sb}
\maketitle


\section{Introduction}
\label{sec:introduction}

Weak gravitational lensing---a powerful probe of the dark universe---is facing many challenges.
One of them is the nonlinear evolution of the large scale structure.
The non-Gaussianity induced by the nonlinear evolution pushes the cosmological information into high-order statistics,
 leading to the loss of constraining power from the 2-point statistics
 (the power spectra and equivalently the correlation functions) alone.
Recently several works have focused on nonlinear local transformation of the 2D lensing and 3D matter and galaxy fields
 to reduce the non-Gaussianity \cite{Neyrinck09,Neyrinck10,Neyrinck11,Joachimi11b,Seo10,Seo11,Yu11}.
This procedure is shown to be promising.
The nonlinear transformation changes the information distribution in the hierarchical statistics.
It makes the cosmic fields close to Gaussian and hence significantly improves the information we can extract from low-order statistics.

Most of these works adopt the logarithmic transformation.
Instead, we choose a Gaussian transformation, which is defined
 as a local monotonic transformation to Gaussianize the one-point PDF \cite{Yu11}.
Throughout the paper we refer to the application of this transformation as the {\it Gaussianization}.
This approach is motivated by the Gaussian copula hypothesis \cite{Scherrer10}.
If this hypothesis is valid, the Gaussianization will make all the statistics Gaussian, in addition to the one-point PDF.

Through N-body simulations, we quantified the performance of the Gaussianization
 on a 2D lensing field by various measures of non-Gaussianity,
 such as skewness, kurtosis, the 5th- and 6th-order cumulants as functions of smoothing radius, and the bispectrum.
We found that the Gaussianization works surprisingly well.
It suppresses the above non-Gaussianity measures by orders of magnitude and effectively reduces them to zero.
This implies that in many exercises we can treat the new field as Gaussian.
The Gaussianization procedure then transports cosmological information into the power spectrum of the new one.
For this reason, theoretical modeling of the weak lensing statistics can be significantly simplified.

However, this does not necessarily mean that the Gaussianization procedure works for real data,
 in which there are various measurement noises.
A prominent one is the shape measurement noise from the random galaxy shapes.
Its impact on the Gaussianization can be figured out straightforwardly in two limits:
(i) When the shape noise is negligible with respect to the lensing signal,
 the Gaussianization works rather well, as shown in paper I.
(ii) When the shape noise overwhelms the lensing signal, the Gaussianization collapses.
Let us look at the case of a sufficiently large pixel with $N\gg 1$ galaxies.
Due to the central limit theorem, the shape noise distribution approaches a Gaussian, as does the one-point PDF.
Hence the Gaussian transformation becomes trivial and the Gaussianization fails.

Hence, whether or not the Gaussianization is applicable to lensing surveys is a nontrivial question.
To investigate the applicability, we add shape measurement noise to the simulated lensing convergence maps.
Following paper I, we then test the performance of the Gaussianization through several measures of non-Gaussianity,
 including the cumulants up to 6th-order of the smoothed field, and the bispectrum.

We find that the performance of the Gaussianization is strongly affected by shape measurement noise.
However, if we Wiener filter the lensing maps first and then perform the Gaussianization,
 it can still work rather well for DES-like or LSST-like surveys.

Our paper is organized as follows: In Sec.\ref{sec:noisemap} we briefly introduce
 the way to construct noisy weak lensing convergence fields using N-body simulation.
Measures of non-Gaussianity, i.e., the measures of the Gaussianization
 performance in the noisy case are presented in Sec.\ref{sec:noise}.
To reduce the degradation of the Gaussianization method caused by shape
measurement noise, we propose
the Wiener filter as a remedy and redo the analysis in Sec.\ref{sec:wiener}.
In Sec.\ref{sec:conclusion} we summarize the results and outline the key issues
for further investigation.


\bfig{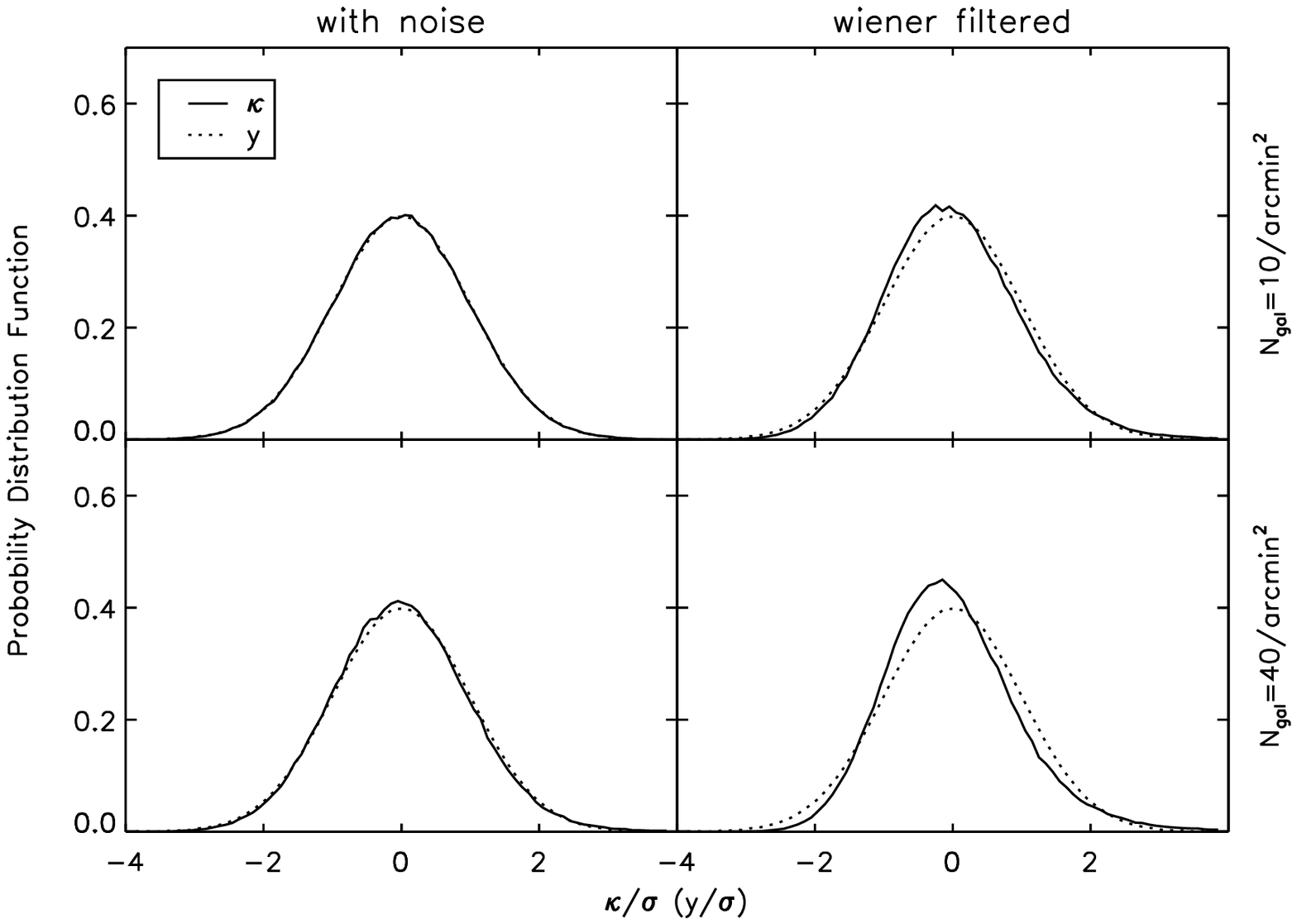}
\caption{The left column shows the PDFs of one convergence map we constructed with two levels of noise added.
The right column shows the PDFs of the convergence map with noise reduction by the Wiener filter.
The PDFs of the field after the Gaussianization are also plotted (dotted line), which is just the standard Gaussian distribution by definition.
From the PDFs we can see that the Gaussian noises of both levels dominate the distribution.
After the Wiener filter is applied, the non-Gaussianity is recovered to some extent.
\label{fig:pdfall}}
\efig

\section{Simulated Lensing convergence maps with shape measurement noise}
\label{sec:noisemap}

We refer the readers to paper I (Yu \etal \cite{Yu11}) and the
references therein for weak lensing basics.
In the first step, we construct noise-free lensing convergence maps, as in paper I.
Our N-body simulation was run using the Gadget-2 code \cite{Gadget2},
adopting the standard $\Lambda$CDM cosmology,
 with $\Omega_m=0.266$, $\Omega_{\Lambda}=1-\Omega_m$, $\sigma_8=0.801$,
 $h=0.71$ and $n_s=0.963$.
The output redshifts are specified such that any two adjacent outputs are separated by the box size $L=300\mpch$ in comoving distance.
More details can be found in \cite{Cui10}.
We stack eight randomly shifted and rotated snapshots to comoving distance $2400\mpch$ to make $\kappa$ maps.
These lensing maps correspond to source galaxy redshift $z_s\approx 1.02$,
 with map size being $7.64^{\circ}\times 7.64^{\circ}$.
When analyzing these maps, we use $512^2$ uniform grids, corresponding to a pixel
size of $0.9^{'}$.  At a typical lens redshift $z_L=0.5$ of a typical
  source redshift $z_s=1$, this corresponds to a size of $0.4\mpch$, well into
  the nonlinear regime. This pixel size is suitable for investigating the
  nonlinear and non-Gaussian regime and for quantifying the performance of
  Gaussianization.  We caution that the Gaussian transformation is nonlinear
  and its dependence on pixel size is nontrivial.  We leave this issue for
  future work.

In observation, the lensing convergence map constructed from cosmic shear
suffers galaxy shape measurement noise,
 with an rms of $\sigma_N\simeq 0.2/\sqrt{N}$, where $N$ is the mean number of
 galaxies per pixel.  In the limit $N\rightarrow \infty$,  the shape noise
 distribution becomes Gaussian according to the central limit theorem. But for
 typical lensing surveys such as DES and LSST, $N\sim 10$-$40$ and the shape
 noise may not fully reach a Gaussian distribution.  Nevertheless, full investigation of the
 form that shape noise takes is well outside the scope of this paper.  To
 proceed,   we approximate the shape noise as Gaussian white noise for
 testing the Gaussianization method.
We add two levels of Gaussian white noise into the convergence maps constructed from
simulation, which correspond to source galaxy number densities of
$10/\mathrm{arcmin}^2$ and $40/\mathrm{arcmin}^2$.  Hereafter we denote the
two cases as case A and case B. The first corresponds to DES-like surveys
 \footnote{Dark Energy Survey, http://www.darkenergysurvey.org} 
and the second corresponds to LSST-like surveys \footnote{Large Synoptic Survey Telescope, http://www.lsst.org}.
In reality, these surveys not only have different source galaxy
number densities, but also different source galaxy redshift distributions
and hence they have different lensing signals. Since our purpose is to investigate
the dependence of the Gaussianization on measurement noise, we neglect the
difference in source redshift distribution and fix the lensing
signal, but vary the measurement noise.  For each noise
level,  we produce 20 Gaussian white noise maps and add them into
convergence maps pixel by pixel.  Throughout the paper, we denote the true
lensing signal as $\kappa_T$, the noise as $\kappa_N$ and the sum as
$\kappa\equiv \kappa_T+\kappa_N$.

For the given pixel size, the corresponding noise rms is $\sim 0.07$ for case
A and $\sim 0.035$ for case B. The lensing signal rms is $\sim 0.02$.
Hence, for both cases the noise dominates over the signal and significantly changes the one-point
PDF, and pushes it towards a Gaussian form.  The numerical results of one
convergence map with two levels of noise added
 are presented in the left column of Fig.\ref{fig:pdfall}. Indeed, for case A,
 there is almost no visible deviation in
 the one-point PDF from the Gaussian distribution. For case B, only small deviations can be
 seen.

\begin{figure*}
\epsfxsize=8cm
\epsffile{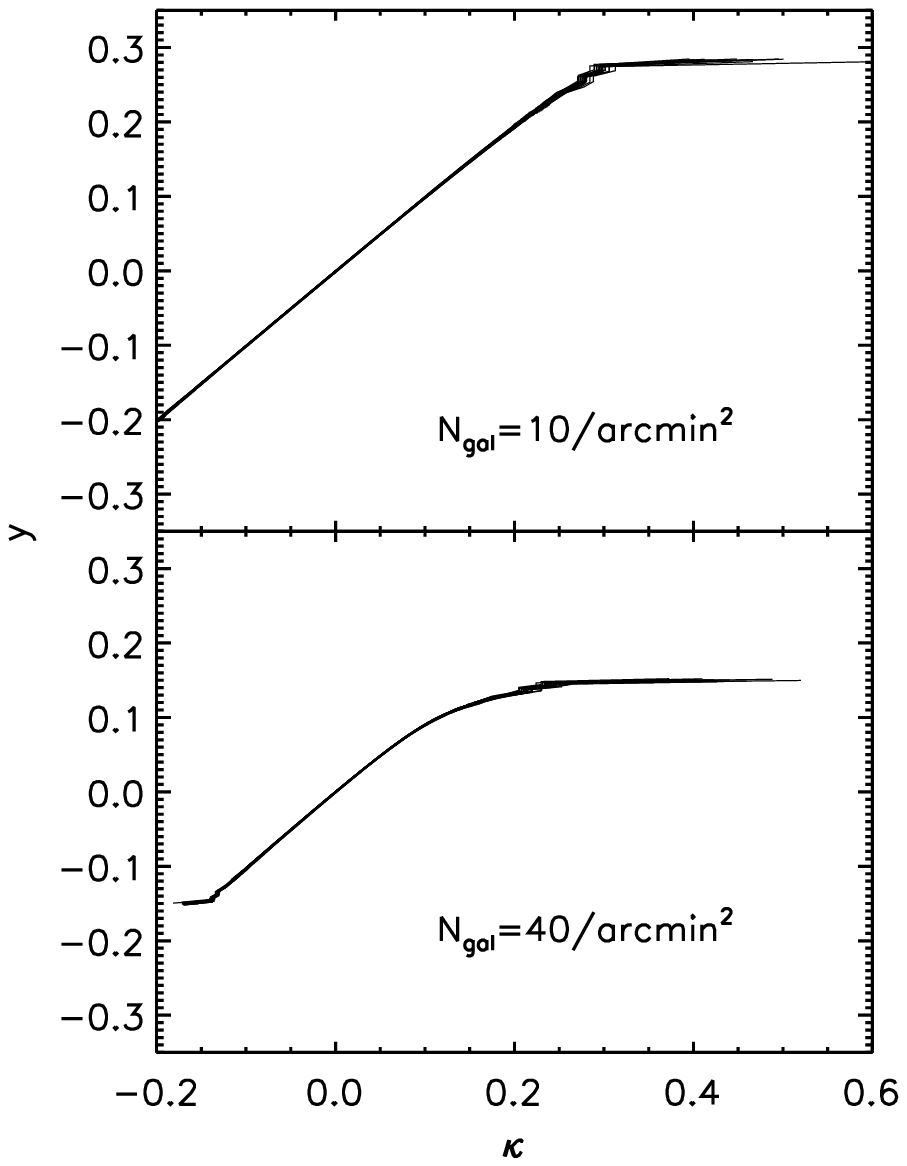}
\epsfxsize=8cm
\epsffile{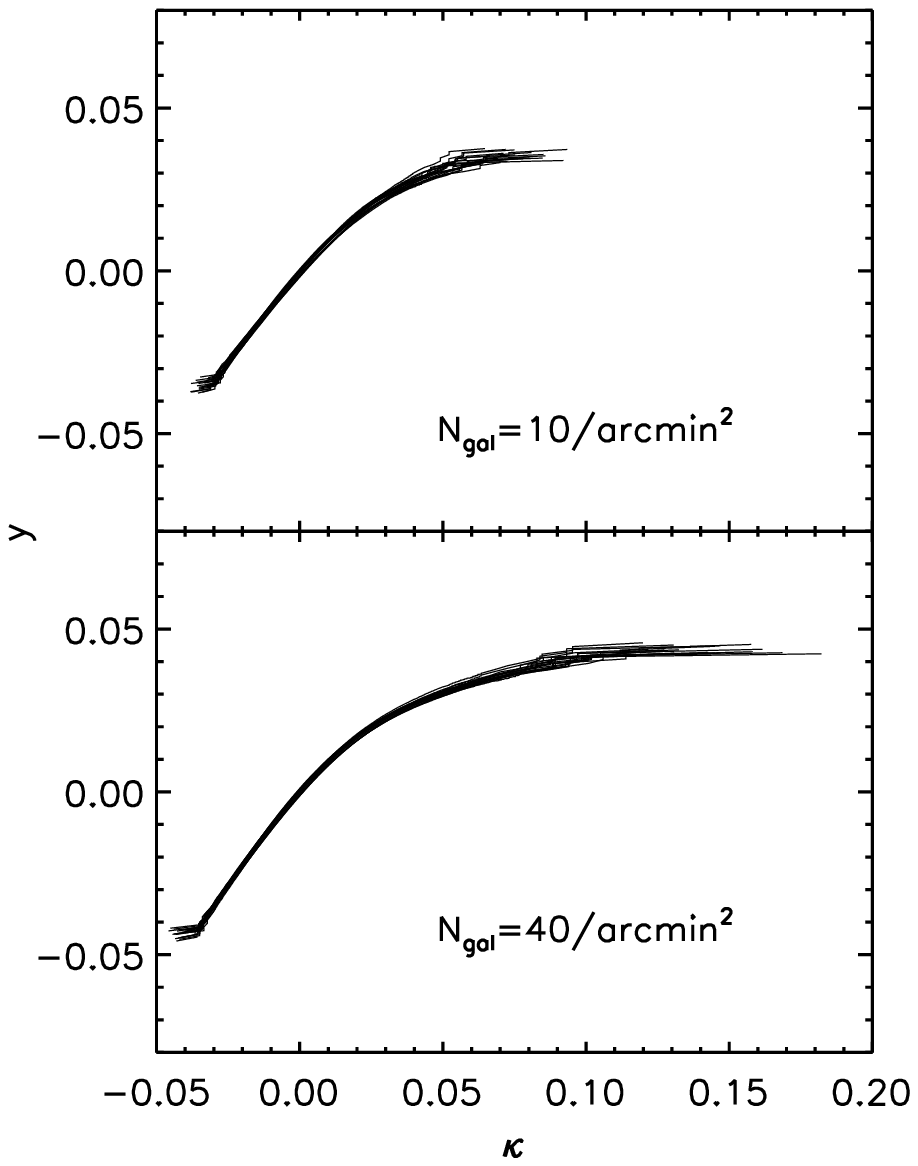}
\caption{The left panel shows the Gaussian transformations from the noisy $\kappa$ to Gaussian $y$ field, for 20 realizations.
The right panel shows the result after the noise is reduced by the Wiener filter.
The heavy noise makes the Gaussian transformations of trivial-mapping ($y=\kappa$).
Only a few pixels survive in case A.
For the noise reduced case, we can see the effect of Wiener filter from the shape of the Gaussian transformation.
 The 20 Gaussian transformations converge with each other for noisy maps,
 and are in reasonable agreement with each other for the noise-reduced case.
The small divergence in the latter case arises from the different normalization factors determined at the zero point.
Nevertheless, the normalization factor will not influence the non-Gaussianity measures.
\label{fig:xxrltnall}}
\end{figure*}
\section{Gaussianization performance on noisy maps}
\label{sec:noise}
Here we briefly introduce the Gaussianization we use. Details are given in
paper I. The Gaussian transformation $\kappa \rightarrow y$ is defined as follows:
\be
\label{eqn:kappay}
\int_{-\infty}^{y}P_G(y){\dif}y=\int_{-\infty}^{\kappa}P(\kappa){\dif}\kappa\ ,
\ee
with normalization at the zero point,
\be
\label{eqn:kappaynormalization}
\left.\frac{{\dif}y}{{\dif}\kappa}\right|_{\kappa=0}=1\ .
\ee
Here by definition $P_G(y)=\exp(-y^2/2\sigma_y^2)/\sqrt{2\pi}\sigma_y$ is the
Gaussian PDF and $\sigma_y$ is the rms dispersion of $y$.
We obtain this transformation numerically for each pixelized simulated map.
The normalization coefficient of the $\kappa\rightarrow y$ relation does not
influence our measures of non-Gaussianity defined below, nor does it alter the performance of Gaussianization.
The particular normalization both in paper I and here (Eq. \ref{eqn:kappaynormalization}) is chosen such that
(i) in the limit that the $\kappa$ field becomes Gaussian, the Gaussianization
transformation becomes an unity transformation,
 and (ii) it shares the same unity slope as the logarithmic transformation at $\kappa=0$.
\footnote{We get transformation $\kappa$ to $y$ with unity $y$ variance first,
 and estimate the slope at $\kappa=0$ by two nearby points.
The normalization coefficient is just the inverse of the slope,
 which makes the transformation normalized but the variance of $y$ non-unity.}

By definition, the one-point PDF of the $y$ field of the given pixel size is
 Gaussian. However,  this does not necessarily mean that the resulting field
 $y$ is a multivariate Gaussian random field.  For example, if we smooth the
 field, the one-point PDF of the smoothed field $y_S$ can be non-Gaussian. Following
 paper I, we will evaluate various non-Gaussianity statistics, such as the
 cumulants as a function of smoothing scale and the bispectrum.  The cumulants
 $K_n$ ($n=1,2,\cdots$) are defined by
\ba
\label{eqn:cumulants}
K_3&\equiv& \frac{\langle y_S^3\rangle}{\langle y_S^2\rangle^{3/2}}\no \ ,\\
K_4&\equiv& \frac{\langle y_S^4\rangle}{\langle y_S^2\rangle^{2}}-3\no \ ,\\
K_5&\equiv& \frac{\langle y_S^5\rangle}{\langle
  y_S^2\rangle^{5/2}}-10\frac{\langle y_S^3\rangle}{\langle
  y_S^2\rangle^{3/2}}\ ,\\
K_6&\equiv& \frac{\langle y_S^6\rangle}{\langle
  y_S^2\rangle^{3}}-10\frac{\langle y_S^3\rangle^2}{\langle
  y_S^2\rangle^{3}}-15 \frac{\langle y_S^4\rangle}{\langle
  y_S^2\rangle^{2}}+30 \no\  .
\ea
The reduced bispectrum is defined by
\be
q(\vec{\ell}_1,\vec{\ell}_2,\vec{\ell}_3)=\frac{B(\vec{\ell}_1,\vec{\ell}_2,\vec{\ell}_3)}{\left[P(\ell_1)P(\ell_2)+{\rm
    cyc.}\right]^{3/4}}\ \ ,
\ee
with $\vec{\ell}_1+\vec{\ell}_2+\vec{\ell}_3=0$.
We express $q$ in the coordinate $(l,u,\alpha)$, where $l\equiv l_1$, $u\equiv
l_2/l_1$ and $ \alpha=\arccos(\vec{l}_1\cdot\vec{l}_2)/l_1 l_2$.

Again, we caution the readers that the definitions of the cumulants and
the reduced bispectrum are different from the widely used definitions in large
scale structure study. They are designed to be invariant under the
transformation $y\rightarrow Ay$, where $A$ is an arbitrary constant. This
property is desirable when comparing these non-Gaussianity measures before and
after the Gaussianization.

\subsection{The $\kappa$-$y$ relation of noisy maps}
\label{subsec:kappay}

The Gaussian transformations we obtain are presented in the left panel of
Fig.\ref{fig:xxrltnall}.  For case A, the noise dominates
over the signal ($|\kappa_N|\geq |\kappa_T|$) for almost all of the pixels. Since the
noise PDF is Gaussian to a good approximation, we  have $y=\kappa$. This is
what we see from  Fig. \ref{fig:xxrltnall}.  On the other hand, visible
deviation from $y=\kappa$ can be found at the high $\kappa$ end, where the
lensing signal can dominate over the noise due to its skewed PDF.

For case B, noise is reduced by a factor of 2, so the non-Gaussianity of the
PDF is less affected. As a consequence, we find larger deviation from
$y=\kappa$.

The Gaussian transformations for 20 realizations converge with each
other very well. The convergence is much better than what we found in the previous
work, in which the lensing maps are free of noise. This better convergence is also
a manifestation of overwhelming shape measurement noise.

\subsection{The cumulants of smoothed noisy maps}
\label{subsec:kappacumu}

\bfig{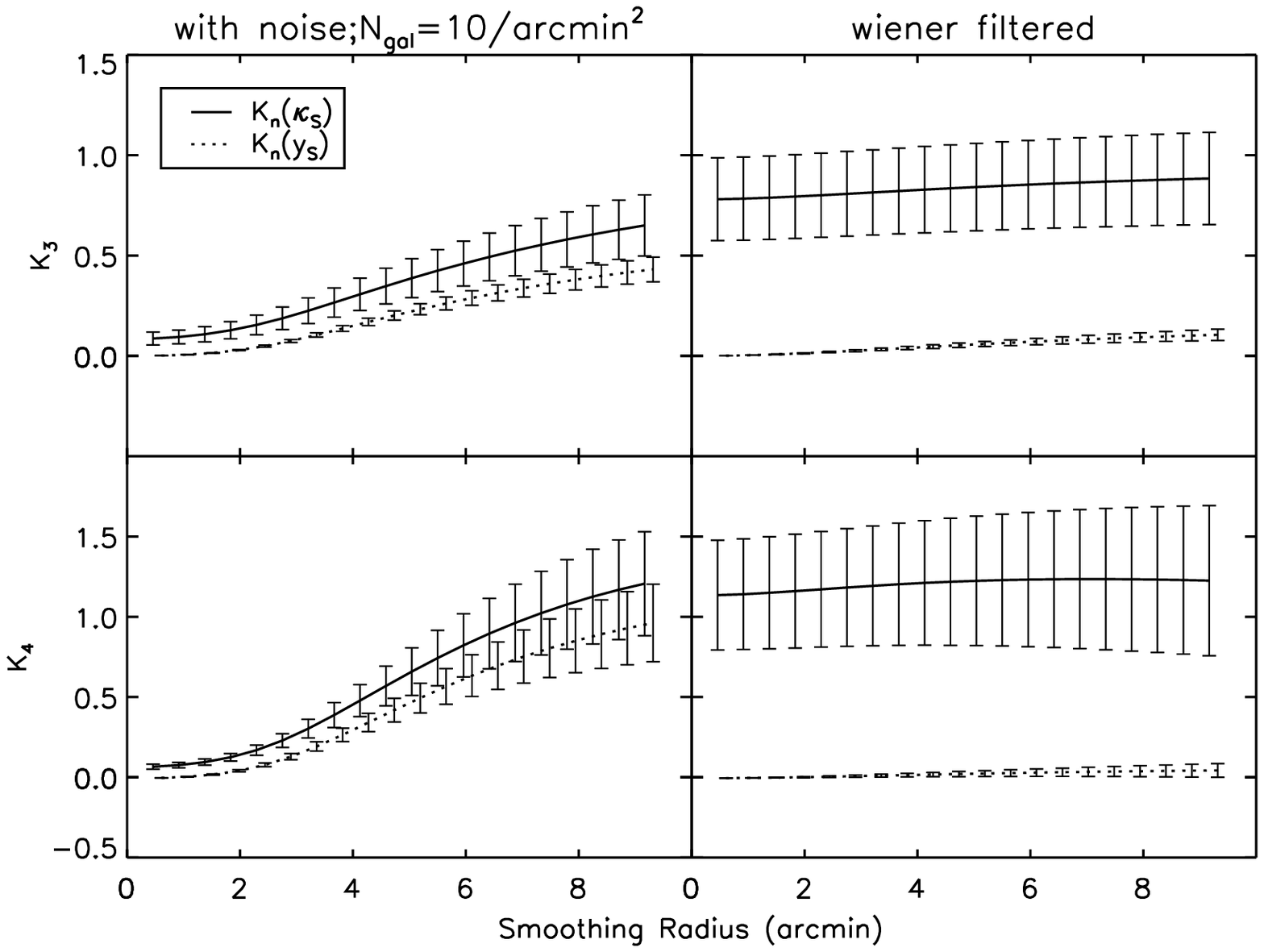}
\caption{The ensemble averaged skewness (top) and kurtosis (bottom) of the smoothed
  noisy (left) and the Wiener filtered (right)
 convergence maps are plotted in a solid line as a function of smoothing scale for case A.
The results of the Gaussianized fields are plotted in a dotted line with a
little horizontal shift for clarity. The error bars are the rms dispersions
among 20 realizations. The Gaussianization cannot reduce the skewness and
kurtosis of these noisy maps to an acceptable level. However, we find that the
Wiener filter is efficient in reducing the shape measurement noise. For the
Wiener filtered maps, the Gaussianization can suppress the skewness and kurtosis
 towards zero for all the smoothing scales we considered.  The
   statistical errors are quite
  correlated, since $K_n$ at different smoothing scales share much the same
  cosmic volume, and hence much the same cosmic variance and measurement
  noise.
\label{fig:s3r01034}} \efig

The cumulants of the smoothed filed $K_n(y_S)$ are nontrivial checks of
non-Gaussianity,  despite the fact that the cumulants of the unsmoothed field $K_n(y)$
are Gaussian by definition.  Smoothing introduces pixel-pixel correlation and
gives rise to non-Gaussian $y_S$ cumulants.
The results for the cumulants of the smoothed field up to the 6th-order are presented in
 the left column of Figs.\ref{fig:s3r01034} and \ref{fig:s3r01056} for case A,
 and in Figs.\ref{fig:s3r04034} and \ref{fig:s3r04056} for case B.
As in paper I, all results are averaged over 20 realizations and
 the error bars are the rms dispersions among these realizations.
 Note that these error bars on different smoothing scales are highly correlated with each other.
To highlight the performance of the Gaussianization, $K_n(y_S)/K_n(\kappa_S)$ as functions of smoothing scale are
 shown in Fig.\ref{fig:s3rcmp}.

Contamination of shape measurement noise significantly changes the behavior of the
cumulants in two ways: (i) Since the noise
field is Gaussian and the measurement noise often dominates the lensing
signal, the cumulants $K_n(\kappa_S)$ are much lower than the corresponding
ones in the noise-free case (paper I).  (ii) Smoothing suppresses the noise
more efficiently than the signal, since the power of noise is more
concentrated to small scales (e.g., zero-lag). Hence for sufficiently large
smoothing scales, the signal can dominate over the noise and the non-Gaussianity of
the signal can show up. This leads to larger $K_n$ for larger smoothing
scales for case A.  On the other hand, smoothing decreases the
non-Gaussianity of the signal.  So for the noise-free case in paper I, $K_n$
decreases with the smoothing scale. Case B has a noise level between case
A and the noise-free case. The two effects of smoothing compete and cause
$K_{4,5,6}$ to increase with the smoothing scale first and then decrease with
it.

The heavy shape measurement noise contamination significantly degrades the
performance of the Gaussianization.  For a shallower lensing survey like DES (case
A), the Gaussianization has very limited capability to reduce non-Gaussianity
(Figs.\ref{fig:s3r01034}, \ref{fig:s3r01056} and \ref{fig:s3rcmp}). As we
expect, the Gaussianization works better for the case of lower noise. But even
for deep surveys like LSST (case B), the Gaussianization can only moderately
suppress the non-Gaussianity. It suppresses the skewness by a factor less than 2.
Although the situation is better for the higher-order cumulants, the suppression
factor is still less than 10 for $K_{4,5,6}$.

\bfig{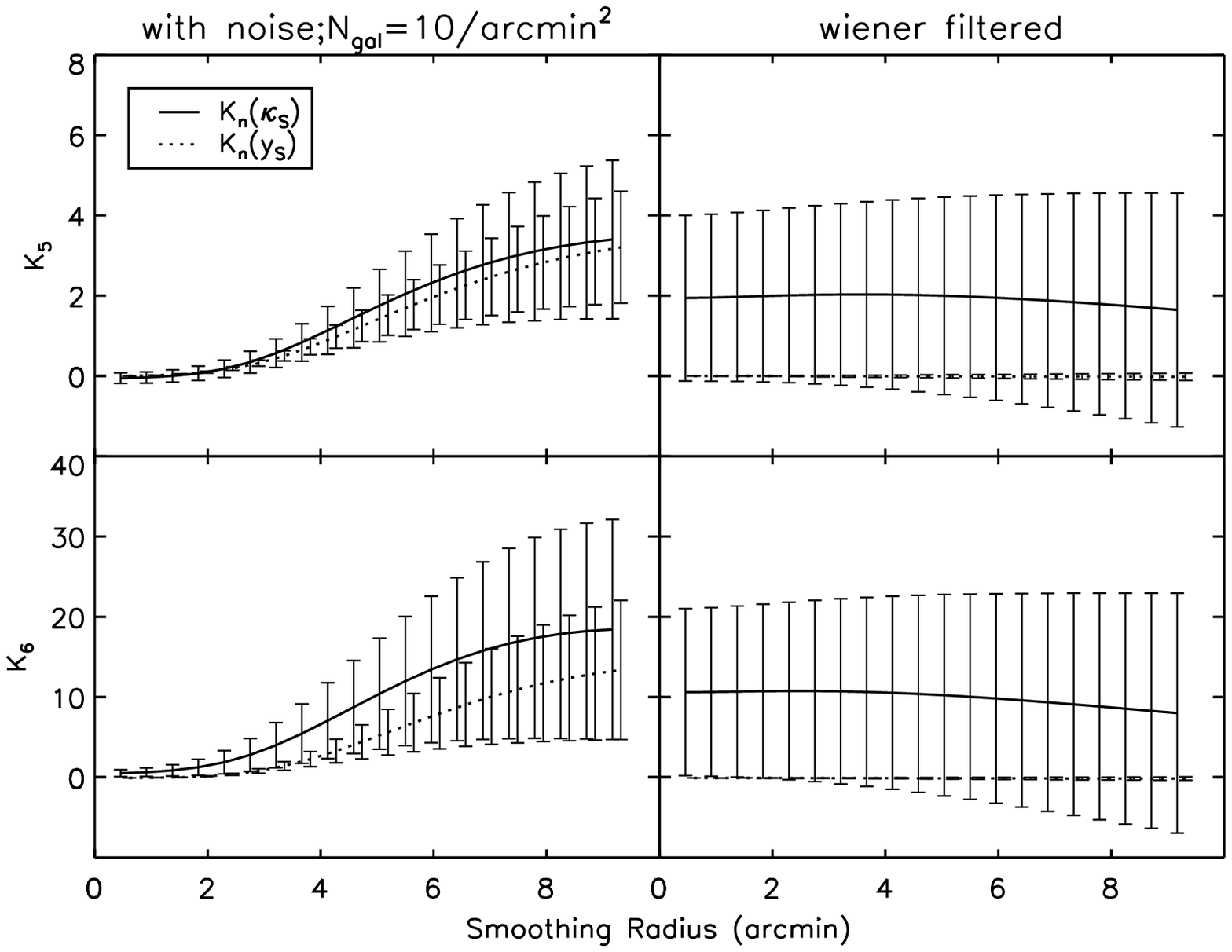}
\caption{The ensemble averaged 5th- (top) and 6th- (bottom) order cumulants of the smoothed noisy (left) and noise-reduced (right)
 convergence maps are plotted in a solid line as a function of smoothing scale for case A.
The results of the Gaussianized fields are plotted in a dotted line with a little horizontal shift for clarity.
The correlated error bars are the rms dispersions among the 20 realizations.
The Gaussianization cannot reduce the 5th- and 6th-order cumulants of these noisy maps to an acceptable level.
When the noise is reduced by the Wiener filter, the Gaussianization can suppress the 5th- and 6th-order cumulants
 to zero for all the smoothing scales we considered.
\label{fig:s3r01056}} \efig

\bfig{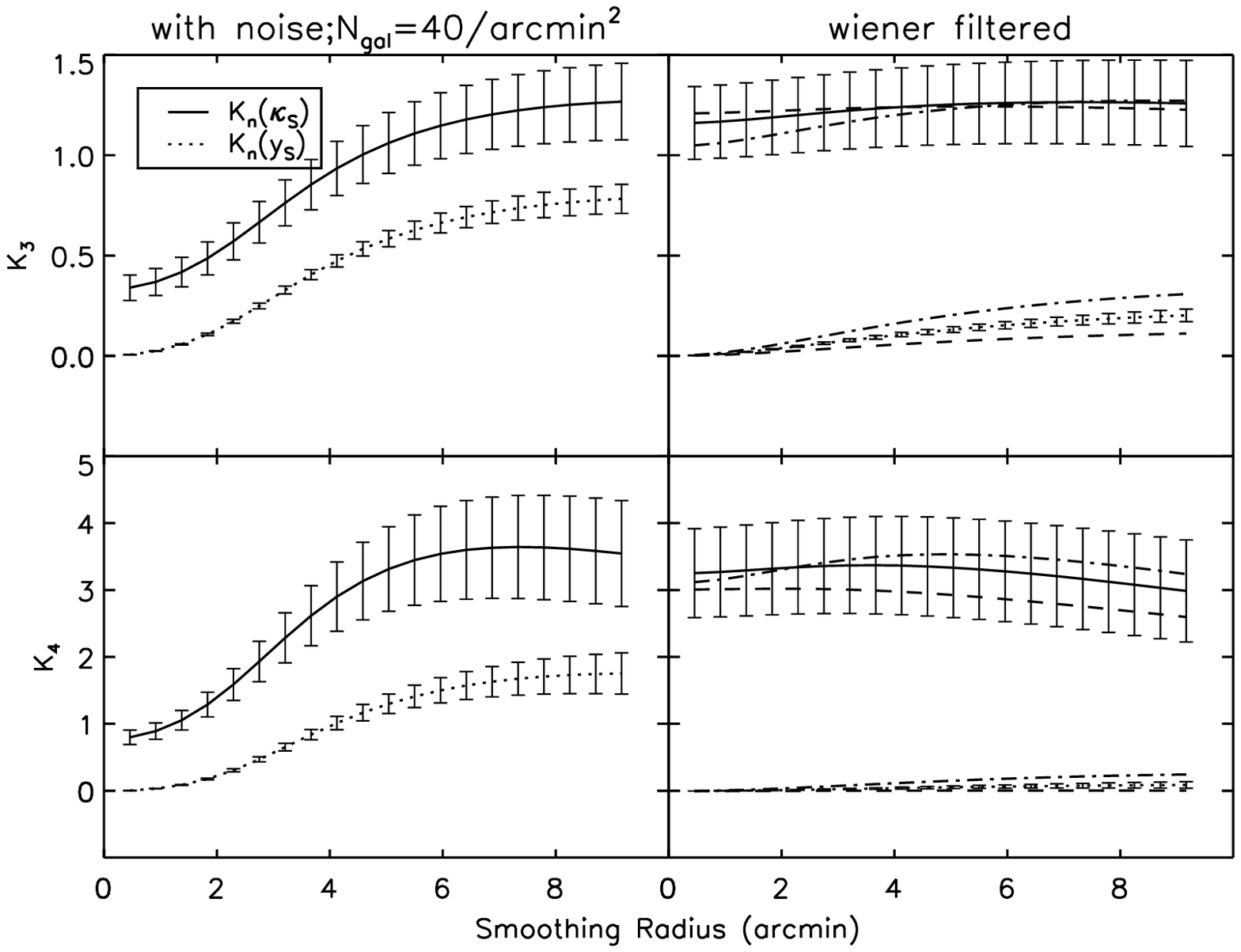}
\caption{The ensemble averaged skewness (top) and kurtosis (bottom) of the smoothed noisy (left) and noise-reduced (right)
 convergence maps are plotted in a solid line as a function of smoothing scale for case B.
The results of the Gaussianized fields are plotted in a dotted line.
The correlated error bars are the rms dispersions among the 20 realizations.
The Gaussianization cannot reduce the skewness and kurtosis of these noisy maps to an acceptable level.
When the noise is reduced by the Wiener filter, the Gaussianization can suppress the kurtosis
 towards zero for all the smoothing scales we considered.
But for the skewness, we still see residual non-Gaussianity.
 The results for an inaccurate noise spectrum estimation used in the Wiener filtering are presented as dashed and dot-dashed lines.
A factor of 2 overestimation or underestimation does not change the Gaussianization performance much.
\label{fig:s3r04034}} \efig

\bfig{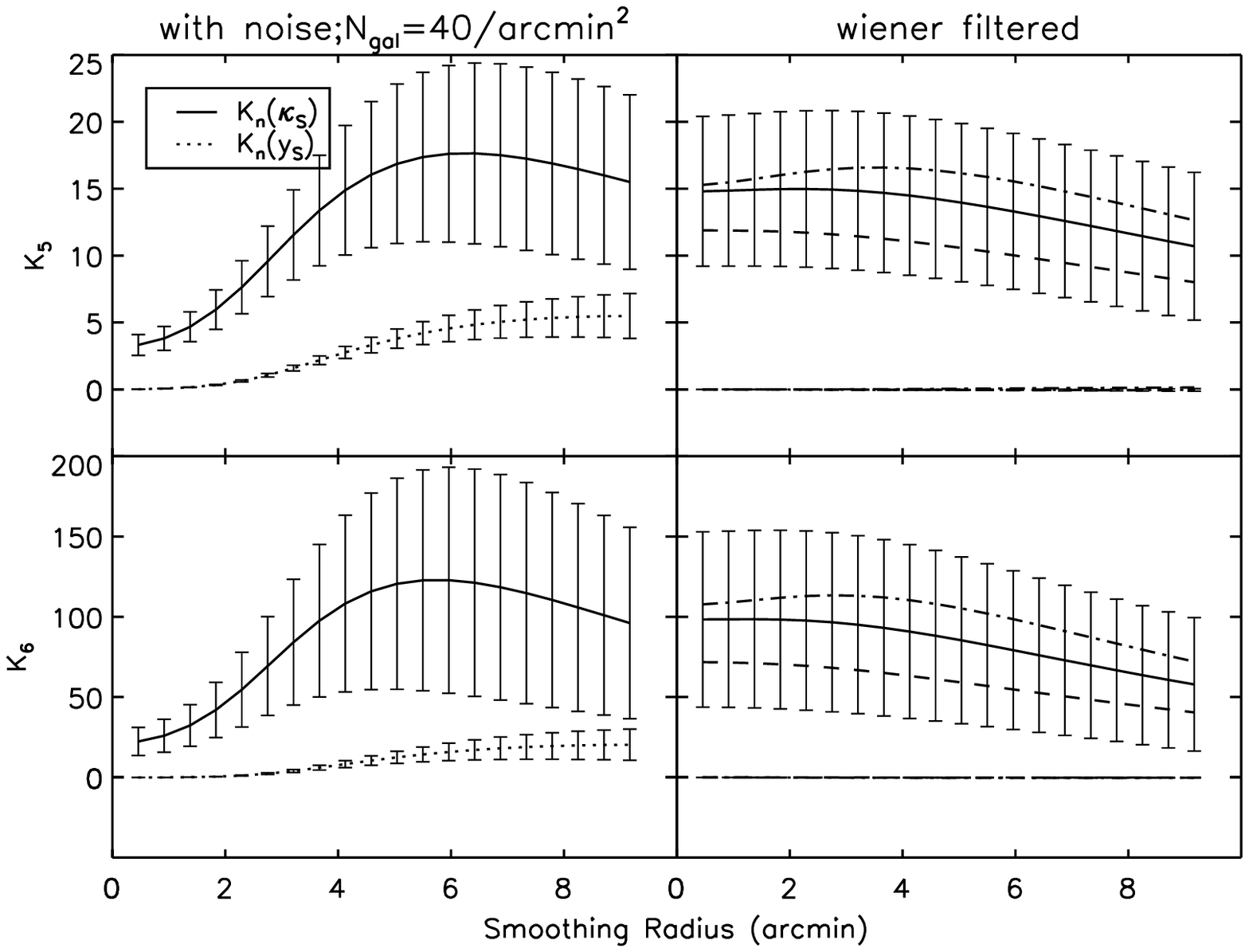}
\caption{The ensemble averaged 5th- (top) and 6th- (bottom) order cumulants of the smoothed noisy (left) and noise-reduced (right)
 convergence maps are plotted in a solid line, as a function of smoothing scale for case B.
The results of the Gaussianized fields are plotted in a dotted line.
The correlated error bars are the rms dispersions among the 20 realizations.
The Gaussianization cannot reduce the 5th- and 6th-order cumulants of these noisy maps to an acceptable level.
When the noise is reduced by the Wiener filter, the Gaussianization can suppress the 5th- and 6th-order cumulants
 to zero for all the smoothing scales we considered.
The results for an inaccurate noise spectrum estimation used in the Wiener filtering are presented as dashed and dot-dashed lines.
A factor of 2 overestimation or underestimation does not change the Gaussianization performance much.
\label{fig:s3r04056}} \efig

\bfig{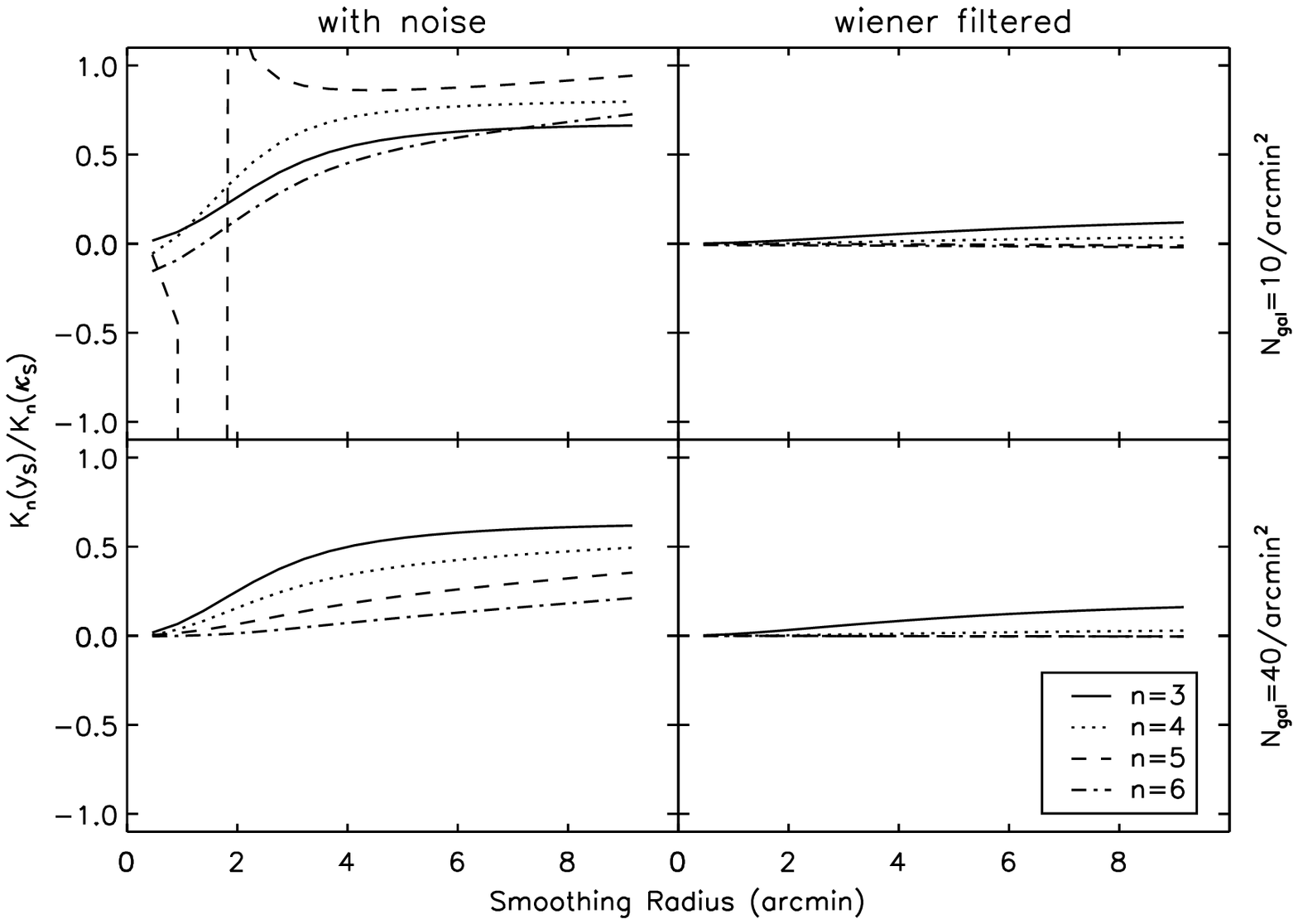}
\caption{The ratios of the $n$th-order cumulants of the smoothed noisy (left) and noise-reduced (right)
 convergence maps from prior to the Gaussianization to afterwards, as functions of smoothing scale
 for case A (top) and B (bottom).
The Gaussianization can only suppress by a factor of 2 for the skewness of a noisy map, and a little more or less for higher order cumulants.
When the noise is reduced by the Wiener filter, the Gaussianization can suppress by a factor of 5 for the skewness of a noisy map,
 and more than 10 for higher-order cumulants.
\label{fig:s3rcmp}} \efig

\subsection{The bispectrum of noisy maps}
\label{subsec:kappabisp}

\begin{figure*}
\epsfxsize=16cm
\epsffile{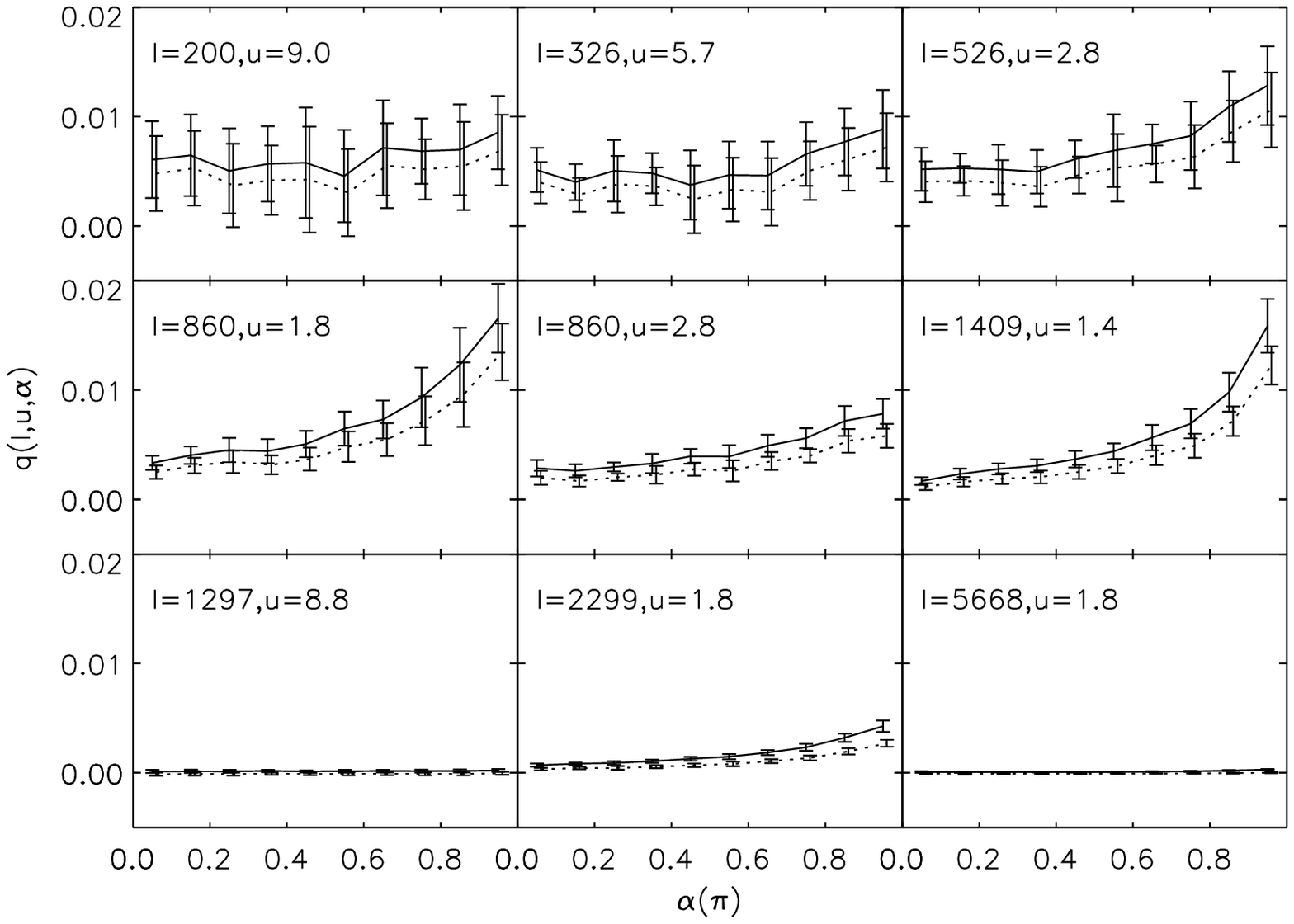}
\caption{The ensemble averaged reduced bispectrum $q(\ell,u,\alpha)$ of the noisy convergence maps for case B
 on the same 9 configurations as the previous work.
The $q$s prior to the Gaussianization are presented in a solid line, while the afterwards is in a dotted line with a horizontal shift for clarity.
The error bars are the rms dispersions over 20 respective maps.
The Gaussianization can only suppress the bispectrum by a small factor,
 and a large amount of non-Gaussianity still remains in bispectrum of the new field.}
\label{fig:nsrdbps}
\end{figure*}

Another measure of non-Gaussianity is the reduced bispectrum $q(\vec{l}_1,\vec{l}_2, \vec{l}_3)\equiv q(\ell, u,\alpha)$,
 which vanishes for Gaussian fields.
The reduced bispectrum for various configurations are presented in Fig.\ref{fig:nsrdbps}.
The Gaussianization almost entirely fails to reduce the bispectrum for case A, in which the shape measurement noise dominates.
So we only show the results for case B.
To better demonstrate the impact of shape noise, we choose the same $9$ configurations of $l\in [200,6000)$ as in paper I.
For most configurations of $(l, u,\alpha)$, $q$ can only be suppressed by the Gaussianization by a small factor.
For many configurations, the $q$s after the Gaussianization deviate significantly from zero, and stay positive.
This result differs from the noise-free case in paper I,
 in which the $q$'s after the Gaussianization are consistent with and scatter
 about zero.  Since the bispectrum is an unambiguous measure of non-Gaussianity,
 these nonzero results show the failure of the Gaussianization.
This poor performance of the Gaussianization is consistent with the results in the previous section on the skewness,
 which is the bispectrum weighted over all configurations.


\section{Gaussianization performance on Wiener filtered maps}
\label{sec:wiener}
The way that Gaussianization works is that it relatively down-weights large
$\kappa$ pixels (regions)
 which are mostly responsible for the non-Gaussianity.
Hence to improve the performance of Gaussianization for noisy lensing maps,
 we shall apply a filter to down-weight regions of low S/N, so regions of
   high S/N can be highlighted. In this way the power of Gaussianization can
   still persist. For this purpose, we will adopt the Wiener filter. The
   Wiener filter is a filter in Fourier space. But
   statistically speaking, it has the desirable property of highlighting high
   S/N regions.   It  is also widely used in astronomy.
In particular, it is shown by \cite{ZTJ05} that the Wiener filter can improve the reconstruction
 of one-point (true) convergence PDFs from noisy lensing maps.

\subsection{Wiener filter}
In signal processing, the Wiener filter is widely used to reduce the amount of
noise presenting in a signal
 by comparison with an estimation of the desired noiseless signal. For
 simplicity, we briefly review the Wiener filter in 1D space.
Assume that the corrupted signal is $\hat{s}(x)=s(x)+n(x)$, in which
 $s(x)$ is the original signal and $n(x)$ is the noise.
We can convolve $\hat{s}(x)$ with a filter $h(x)$ to reconstruct the signal
$s(x)$. The reconstruction error has  a dispersion
\be
\label{eqn:wferr}
E(H(k))=\int\dif k\left|H(k)[S(k)+N(k)]-S(k)\right|^2,
\ee
in which $H(k)$, $S(k)$, $N(k)$ are the Fourier transforms of $h(x)$, $s(x)$
and $n(x)$, respectively. The Wiener filter is the one to minimize $E$,
\be
\label{eqn:wiener}
H(k)=\frac{|S(k)|^2}{|S(k)|^2+|N(k)|^2}\ .
\ee
At scales where the signal dominates over the noise,
 the filter is close to 1, while at scales where the noise dominates over the
 signal, the filter is close to 0. This is a desirable property for our
 purpose.

In our application, the $|S(k)|^2$ term in the Wiener filter is
 the power spectrum of the $\kappa$ field
 $P(\ell)=\langle|\kappa(\vec{\ell})|^2\rangle$.
 The $|N(k)|^2$ term is the Gaussian white shape noise power spectrum, which
 does not vary with scale.

We will basically redo the analysis from Sec.\ref{sec:noise} for noise reduced
maps by the Wiener filter Eq.(\ref{eqn:wiener}).
 Since in observation the noise power spectrum will not be known exactly,
we also test the Wiener filtering in which the noise power spectrum is overestimated
or underestimated by a factor of 2 for case B. We may expect that the noise
power spectrum estimated in a real survey would fall somewhere between the two
extreme cases.

The PDFs of the Wiener filtered convergence maps are presented in the right column of Fig.\ref{fig:pdfall}.
Compared to the unfiltered maps in the left column, the Wiener filter indeed
recovers the lensing signal to some extent, and as a consequence the PDFs are
now visibly non-Gaussian, even for case A, which has larger measurement
noise. Nevertheless, the Wiener filter significantly improves the PDF
recovery.
Later we will show the gain of non-Gaussianity of the maps.
But compared to the strong non-Gaussianity existing in the noise-free case,
 it is obvious that the Wiener filter can not recover the field too much.
For case A, the PDF is still close to a Gaussian distribution.
The recovery effect is more obvious for case B.

\subsection{The $\kappa$-$y$ relation of noise-reduced maps}
\label{subsec:wfkappay}

The Gaussian transformations for the noise-reduced convergence maps are presented in the right panel of Fig.\ref{fig:xxrltnall}.
Compared to the left panel, the shape of the transformation $\kappa\rightarrow y$ apparently changes.
The turning points between the straight line $y = \kappa$ (which represents
the dominance of Gaussian noise)
 and the high $\kappa$ end (where the Gaussianization takes effect) shift
 towards a low $\kappa$ value.
 Though the reduced noise still makes the Gaussian transformation a trivial mapping for the low $\kappa$ value pixels,
 pixels with higher $\kappa$ persist through the noise thanks to the filtering,
 and thus the Gaussianization method will have effects on these high $\kappa$ value pixels.
The 20 Gaussian transformations after noise reduction are in reasonable agreement with each other for both case A and B.
The small divergence in this noise-reduced case rises from the different normalization factors determined at the zero point.
Nevertheless, the normalization factor will not influence the non-Gaussianity measures.

\subsection{The cumulants of noise-reduced maps}
\label{subsec:wfkappacumu}

The cumulants $K_{3,4,5,6}$ for the noise-reduced case are presented in
 the right column of Figs.\ref{fig:s3r01034} and \ref{fig:s3r01056} for case A
 and Figs.\ref{fig:s3r04034} and \ref{fig:s3r04056} for case B.
The solid lines are for the filtered convergence maps prior to the Gaussianization $K_n(\kappa_S^{\rm WF})$,
 while the dotted lines are for afterwards $K_n(y_S^{\rm WF})$, in which superscript ${\rm WF}$ means that the noise is reduced by the Wiener filter.
All the results are averaged over 20 realizations and the error bars are the rms dispersions.
The ratios of the cumulants  $K_n(y_S^{\rm WF})/K_n(\kappa_S^{\rm WF})$ are presented in the right column of Fig.\ref{fig:s3rcmp}.

The maps with noise reduction have more skewness than the case without noise reduction (left column),
 and have less skewness than the noise-free case in the former work, which is just as we would expect from the PDF result.
On large smoothing scales, the recovery of the non-Gaussianity due to the smoothing out of the noise and
 the reduction of the non-Gaussianity due to the smoothing of the signal, seem to just cancel each other
 for case A, leading to an approximately constant skewness.
Similar results also occur for $K_{4,5,6}$.
In case B, for $K_{4,5,6}$ on large smoothing scales,
the reduction effect dominates over the recovery effect.
We can see the cumulants stay constant first on small smoothing scales due to the equilibrium of the two opposite effects,
 and decrease on large smoothing scales due to the overwhelming reduction effect.

For both case A and B, for all order cumulants we considered,
 the Gaussianization works much better than the case without noise reduction.
But we still see nonzero skewness on the largest smoothing scale.
The kurtosis and the 5th- and 6th-order cumulants of the Gaussianized maps are well consistent with zero within errors.
The Gaussianization also reduces the variance among different realizations.

We can quantify the performance of the Gaussianization in the right column of Fig.\ref{fig:s3rcmp}.
For both case A and B, the Gaussianization can suppress the skewness by a factor of 5
 on the largest smoothing scale, and better (more than 10) in higher-order cumulants.
The Gaussianization has much better performance on the Wiener filtered case (as we expected),
since the Wiener filter reduces the noise and recovers the signal to some extent,
 and the Gaussianization has an effect on more pixels that are dominated by the non-Gaussian signal.

\subsection{The bispectrum of noise-reduced maps}
\label{subsec:wfkappabisp}

\begin{figure*}
\epsfxsize=16cm
\epsffile{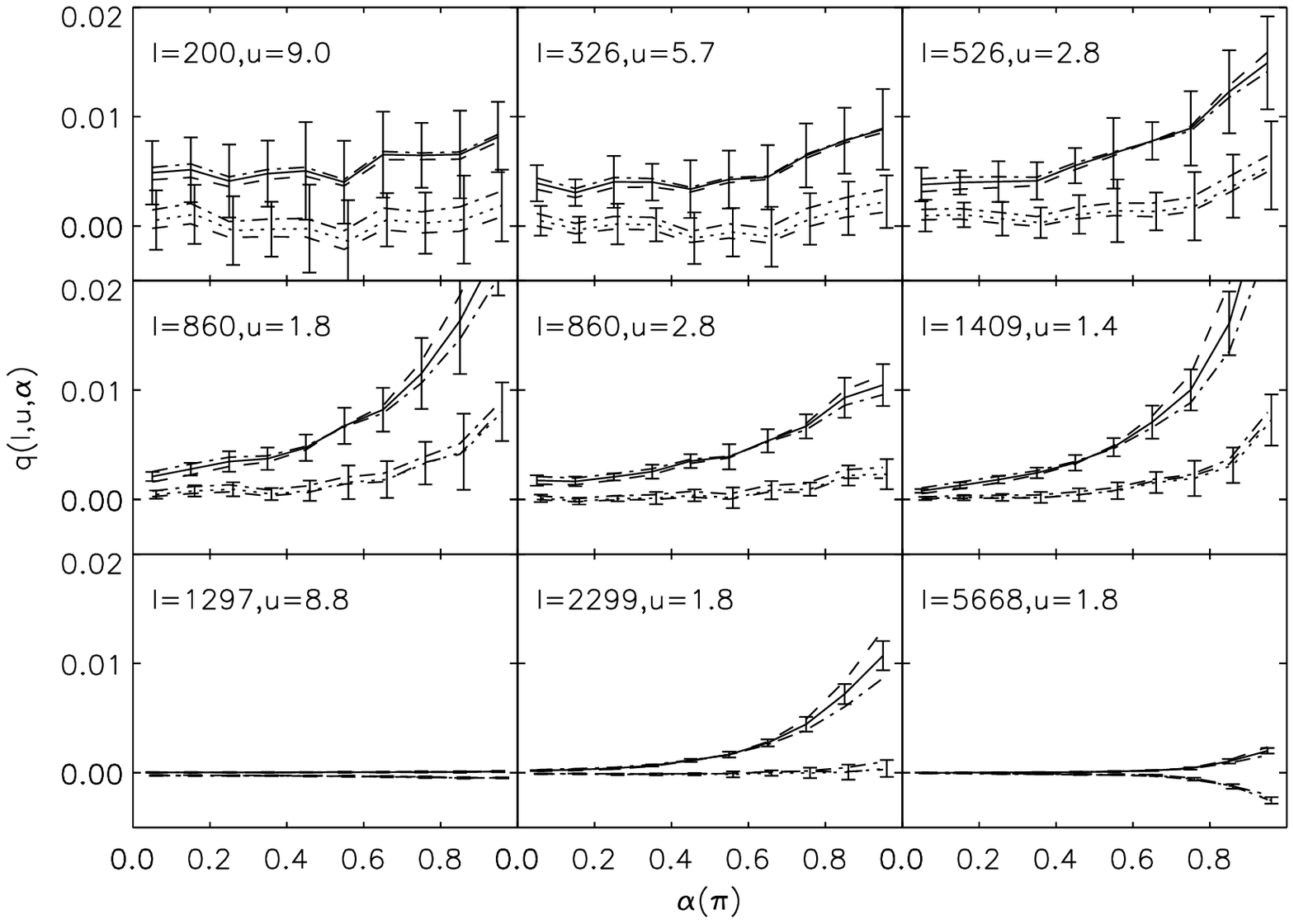}
\caption{The ensemble averaged reduced bispectrum $q(\ell,u,\alpha)$ of the noise-reduced convergence maps for case B,
on the same 9 configurations.
The $q$s prior to the Gaussianization are presented in a solid line and those afterwards in a dotted line with a horizontal shift for clarity.
The error bars are the rms dispersions over 20 respective maps.
The recovery effect of the Wiener filter results in a larger value in the bispectrum.
The Gaussianization after noise reduction can suppress the bispectrum more,
 and on many configurations the bispectrum afterwards is close to zero within error bars.
But there still exists some amount of non-Gaussianity on small scale configurations.
 The results for an inaccurate noise power spectrum estimation used in the Wiener
  filtering are presented in dashed and dot-dashed lines, for cases of
  a factor of 2 overestimation or underestimation in the noise power spectrum.
  The results show that, even for these extreme cases, the
  Gaussianization performance remains impressive.}
\label{fig:wfrdbps}
\end{figure*}

The reduced bispectrum of noise-reduced maps for various configurations are presented in Fig.\ref{fig:wfrdbps},
 where the solid lines are for the noise-reduced fields $q(\kappa^{\rm WF})$ and dotted lines are for the corresponding $q(y^{\rm WF})$ fields.
The results are averaged over the 20 realizations and the error bars are the rms dispersions within them.
We show the bispectrum on the same $9$ configurations as the $q$'s in the last section and in the former work.
For most configurations of $(l, u,\alpha)$, $q(\kappa^{\rm WF})$'s prior to the Gaussianization have larger values
 than the results in the previous section, due to the recovery of non-Gaussianity by the Wiener filter.
We can see that after the noise reduction, for many configurations $q$'s can be significantly suppressed by the Gaussianization,
 and the $q(y^{\rm WF})$'s are consistent with zero within error bars.
These results show the effectiveness of the Gaussianization after the noise reduction by the Wiener filter.
This conclusion is consistent with the findings in the previous section on the skewness.

 Since $q$ describes the non-Gaussianity of an individual configuration,
 to some extent it is a more sensitive measure of the performance of the Gaussianization than skewness.
We find robust evidence for residual non-Gaussianity in some configurations of the bispectrum
 (e.g., some configurations with large $\alpha$ in Fig. \ref{fig:wfrdbps}).
This means that the non-Gaussianity and pixel correlation is also not completely removed by the Gaussianization,
 which is the same situation as in the former work.
A direct implication is that the Wiener filter can not fully recover the copula of the convergence field.
Another source is the small deviation from the Gaussian copula of the noise-free convergence field, as we found in paper I \cite{Yu11}.

\subsection{Inaccurate noise estimation in the Wiener filter}
\label{subsec:wfkappabisp}

We test whether a precisely known noise power spectrum is crucial in
  Wiener filtering. We mimic uncertainties in noise power spectrum
  determination by artificially multiplying the correct noise power spectrum by a
  factor of 2 or $1/2$.
The results are presented in the dashed line and dot-dashed line
 in Figs.\ref{fig:s3r04034}, \ref{fig:s3r04056} and \ref{fig:wfrdbps} for case
 B. The Gaussianization is indeed affected by uncertainties in the noise power
 spectrum determination. However, even for a rather large overestimation or
 underestimation by a factor of $2$,  the Gaussianization remains efficient.
 This behavior is more or less what we expect. The Wiener filter works for the
 Gaussianization since it approaches one where signal overwhelms noise, and
 approaches zero where noise overwhelms signal.  A wrong noise power
 spectrum does not alter this crucial property; it only affects the
 transition region where signal is comparable to noise. As long as the noise
 estimation is reasonably good, the Wiener filter would be robust in improving
 the Gaussianization performance.

Furthermore, we notice an interesting fact: an inaccurate estimation of the
noise spectrum used in the Wiener filter may improve the performance of Gaussianization.
This implies that the Wiener filter may not be  the optimal filter to improve the
Gaussianization. This issue should be investigated further.


\section{Discussions and conclusions}
\label{sec:conclusion}

We have demonstrated that the Gaussianization method we proposed in the former work
 fails when Gaussian shape noise is added into the convergence maps.
We quantify the performance of the Gaussianization by various measures of the non-Gaussianity,
 such as skewness, kurtosis, the 5th- and 6th-order cumulants of the smoothed fields, and the bispectrum.
We found that the Gaussianization can only suppress the skewness by a factor less than 2,
 and more or less for higher-order cumulants for different noise levels.
This implies that we can not treat the resulting $y$ field as Gaussian,
 since higher-order statistics are still large due to the noise corruption.

We propose the Wiener filter to improve the performance of the Gaussianization
for the real noise existing case.
We find that the Gaussianization works well on the Wiener filtered convergence
maps. We also find that the performance is rather robust against uncertainties
in determining the noise power spectrum in the Wiener
filtering. So an accurate determination of the noise power spectrum is not
crucial in applying the Wiener filter and the Gaussianization to observational
data. This insensitivity improves the applicability of the Gaussianization to
real data. The Gaussianization procedure can transport cosmological information in
higher-order statistics  of noisy lensing fields into the power spectrum of $y$.
For this reason, analyzing weak lensing statistics can be significantly simplified.
Nevertheless, measurement noise causes irreducible corruption in the
Gaussianization performance, compared to the noise-free case.

Due to the lack of fully independent realizations, finite box size and limited
number of particles of the simulation,
 our findings are inevitably affected by numerical issues such as
 cosmic variance, mass and force resolution, shot noise and aliasing effect \cite{Jing05}.
We do not expect that any of these factors will alter the major conclusion on the
 failure of Gaussian transformation due to the heavy noise and the success after the noise reduction by the Wiener filter.
These numerical issues could in particular have a larger impact on the residual non-Gaussianity,
 which itself is weak and relatively hard to measure accurately.
However, it is very unlikely that the detected residual non-Gaussianity can be canceled exactly.
In this sense, the detection of the residual non-Gaussianity is robust,
 although we need many more simulations to measure the amplitude to high precision.

 A number of works have addressed extra complexities in Gaussianization.
Neyrinck \etal \cite{Neyrinck10} checked the effect of the logarithmic transformation
 and Gaussian transformation on the galaxy density field.
They found that---although due to the presence of discreteness noise the galaxy density field cannot be fully Gaussianized---
 both transformations still dramatically reduce nonlinearities in the power spectra of cosmological matter and galaxy density fields.
But the transformations do increase the effective shot noise.
Neyrinck \cite{Neyrinck11} studied the sensitivity of five cosmological parameters constrained
 to the Gaussianized power spectra by logarithmic transformation and the Gaussian transformation.
He found the power spectrum of the log-density provides the tightest cosmological parameter error bars in all five parameters tested.
From our studies, we found that for both the noise-free case and the noisy case
 the Gaussian transformation apparently deviates from the logarithmic transformation.
Although the logarithmic transformation will not produce a perfect Gaussian random field,
 there is no free parameter in the transformation.
Thus the information will not be lost into the freedom of the transformation form,
 and theoretical predictions can be made, such as \cite{wangxin11}.

Joachimi \etal \cite{Joachimi11b} employed an extended form of the logarithmic transformation, using Box-Cox transformations
$\tilde\kappa(\lambda,a)=[(\kappa+a)^\lambda-1]/\lambda $
 with two free parameters $(\lambda\neq 0,a)$ to Gaussianize ideal weak lensing convergence field.
(The Box-Cox transformation reduces to the logarithmic transformation when $\lambda=0$.)
The optimized Box-Cox transformation is determined by fitting it to a Gaussian transformation through the free parameters.
They developed analytical models for the transformed power spectrum, including effects of noise and smoothing, and found that
 the optimized Box-Cox transformation performs better than an offset logarithmic transformation in Gaussianizing the convergence;
 but, none of them are capable of eliminating correlations of the power spectra between different angular frequencies.
When they add a realistic level of shape noise, all the transformations perform poorly with little decorrelation of angular frequencies,
 and the arc-tangent logarithmic transformation---which approximates a straight line near the zero point---
 was proposed to deal with the shape noise situation.

Carron \cite{Carron2011} shows a peculiarity of fields with heavy tails:
 the hierarchical statistics (N-point functions) do not entirely specify the distribution.
Thus the cosmological information is not only pushed towards higher-order
statistics due to the nonlinear evolution,
 but can also become inaccessible through the extraction of the
 full series of moments of the field.
His more recent work \cite{Carron2012a} shows that the entire hierarchy of
moments quickly ceases
 to provide a complete description of the convergence 1-point PDF when leaving
 the linear regime.
However, a simple logarithmic mapping makes the moment hierarchy well-suited
again for parameter extraction.
Carron \cite{Carron2012b} shows that the power spectrum of the log-density field
 carries more information than the power spectrum of the field when entering the
 nonlinear regime. These works indicate that the logarithmic
transformation and the Gaussianization method may work beyond simply
propagating cosmological information in the moment hierarchy, and support the
application of these methods in data analysis and cosmology.

Nevertheless, there are many key issues to be explored in future studies.
An incomplete list includes the following.
\bi
\item The applicability to real data.
  Although in this work we consider a more realistic case than in the former work,
   it still differs from real observations.
  The application is complicated by various measurement errors in real data.
  As long as the pixel size is sufficiently large, the central limit theorem drives its distribution to be Gaussian.
  Since the Gaussianization that we have proposed is nonlinear, it will render this Gaussian noise into a non-Gaussian one.
  The Wiener filter could make the situation more complicated.
  Even worse, the same nonlinear transformation and the Wiener filter could mix the lensing signal and measurement noise.
\item The residual non-Gaussianity.
  The proposed Gaussianization works expectedly worse in the noisy case
   and the Wiener filter deals with the noise surprisingly well,
   but it does not perfectly produce a Gaussian random field.
  The residual non-Gaussianity we have detected is weak and is unlikely to carry a significant amount of cosmological information.
  Nevertheless, it may still be worth investigating the cosmological information carried by the residual non-Gaussianities.
\item Other methods dealing with noise corruption.
  The Gaussianization we proposed fails due to the noise corruption, and succeeds in the noise-free case.
  The results of the Wiener filter set in somewhere in between.
  There may exist some other appropriate tool dealing with the noise corruption which
   meanwhile keeps the effectiveness of the Gaussianization.
\ei

\section*{Acknowledgment}
P.J.Z. thanks the support of the national science foundation of China (grant
No. 10821302, 10973027 \& 11025316), the CAS/SAFEA International Partnership
Program for Creative Research Teams (KJCX2-YW-T23)
and the 973 program grant No. 2009CB24901.
W.P.L. acknowledges the supports of the Chinese National 863 project
(No. 2006AA01A125),  NSFC project (10873027, 10821302), and the
Knowledge Innovation
Program of the Chinese Academy of Sciences (grant KJCX2-YW-T05).
W.G.C. acknowledges the support from the European Commission's Framework
Programme 7,
through the Marie Curie Initial Training Network CosmoComp (PITN-GA-2009-238356).
James N. Fry thanks Shanghai astronomical observatory for the hospitality.



\begin{thebibliography}{}

\bibitem{Yu11} Y. Yu, P. Zhang, W. Lin, W. Cui, J. N. Fry,
 \bibprd {\bf 84}, 023523 (2011).

\bibitem{Neyrinck09} M. C. Neyrinck, I. Szapudi, and A. S. Szalay,
  \bibapjl {\bf 698}, L90-L93 (2009).

\bibitem{Neyrinck10} M. C. Neyrinck, I. Szapudi, and A. S. Szalay,
  \bibapj {\bf 731}, 116 (2011).

\bibitem{Neyrinck11} M. C. Neyrinck,
  \bibapj {\bf 742}, 91 (2011).

\bibitem{Joachimi11b} B. Joachimi, A.N. Taylor, and A. Kiessling,
  \bibmnras {\bf 418}, 145 (2011).

\bibitem{Seo10} H.-J. Seo, M. Sato, S. Dodelson, B. Jain, and M. Takada,
  \bibapj {\bf 729}, L11 (2011).

\bibitem{Seo11} H.-J. Seo, M. Sato, M. Takada, and S. Dodelson,
  \bibapj {\bf 748}, 57 (2012).

\bibitem{Scherrer10} R. J. Scherrer, A. A. Berlind, Q. Mao, and C. K. McBride,
  \bibapjl {\bf 708}, L9-L13 (2010).

\bibitem{Gadget2} V. Springel.
  \bibmnras {\bf 364}, 1105 (2005).

\bibitem{Cui10} W. Cui, P. Zhang, and X. Yang.
  \bibprd {\bf 81}, 103528 (2010).

\bibitem{ZTJ05} T. Zhang, U.-L Pen,
  \bibapj {\bf 635}, 821 (2005).

\bibitem{Jing05} Y. P. Jing,
  \bibapj {\bf 620}, 559 (2005).

\bibitem{wangxin11} X. Wang, M. Neyrinck, I. Szapudi, A. Szalay, X. Chen, J. Lesgourgues, A. Riotto, and M. Sloth,
  \bibapj {\bf 735}, 32 (2011).

\bibitem{Carron2011} J. Carron,
  \bibapj {\bf 738}, 86 (2011)

\bibitem{Carron2012a} J. Carron,
  \bibprl {\bf 108}, 071301 (2012).

\bibitem{Carron2012b} J. Carron, M. Neyrinck,
  \bibapj {\bf 750}, 28 (2012).

\end{thebibliography}
\end{document}